\documentclass[final,5p,times,english,twocolumn]{elsarticle}

\usepackage{amssymb}
\usepackage{tensor}

\usepackage{amsmath}
\usepackage{xcolor}
\usepackage{subcaption}
\usepackage{dsfont}
\usepackage{mhchem}
\usepackage{background}

\journal{arXiv}

\usepackage{hyperref}
\hypersetup{
     colorlinks=true,
      linkcolor=blue,
}

\DeclareMathOperator{\arsinh}{arsinh}
\DeclareMathOperator{\erf}{erf}

\backgroundsetup{contents=}
\begin{document}

\begin{frontmatter}



\title{Uniform volume heating of mixed fuels within the
ICF paradigm}


\author{Hartmut Ruhl and Georg Korn}

\address{Marvel Fusion, Theresienh\"ohe 12, 80339 Munich, Germany}

\begin{abstract}
The paper investigates the feasibility of achieving uniform high-power
volume heating for a fusion reactor concept employing a mixed fuel
composition involving $\text{pBDT}$. The realm of mixed fuel fusion
concepts remains relatively unexplored. The pursuit of uniform
high-power volume heating presents a technological
challenge, yet it bears ramifications for fusion reactor
designs. In this study, we introduce the proposition of employing
embedded nano-structures that represent structured foams. These
structured foams interact with short-pulse lasers, thereby achieving
ultra-high power volume heating both within the fuel and the adjacent
hohlraums. Notably, structured foams exhibit superior efficiency
compared to unstructured foams, plasma or surfaces when it comes to
absorbing high-power, short-pulse lasers. The suggested incorporation
of these embedded structured foams interacting with an array of
ultra-short laser pulses offers a high laser absorption power
density, along with meticulous control over energy and power
distribution within the fuel, both in spatial and temporal
dimensions. This holds the potential for the realization of fusion
reactors characterized by straight-forward designs and low complexity,
where $Q_F \approx Q_T > 1$ is expected for the fuel and target
gains. Depending on the fuel composition they can be strong neutron
sources.
\end{abstract}


\end{frontmatter}


\tableofcontents

\section{Introduction}\label{intro}
The indirect drive ICF approach to nuclear fusion has recently achieved a
milestone at LLNL \cite{banks2021significant,Zylstra2022,
Abu-Shawareb2022,Kritcher2022,Zylstra2022b,DOE2022,
LLNL2022,NYT2022,Science2022} demonstrating the principal viability
of inertial confinement fusion for energy production. However,
the implementation of the ICF concept at LLNL has limitations, as
highlighted in \cite{hurricane2022}. This raises the question of
whether there are alternative approaches to nuclear fusion that could
be commercially viable.

The paper examines the feasibility of a volume-heated mixed fuel 
reactor using $\ce{pBDT}$ with a fusion gain $Q_F \approx Q_T >
1$. Mixed fuel fusion reactor concepts are relatively new, and their
potential effectiveness is still uncertain.

In this paper, we propose a novel direct drive fast heating
concept that relies on embedded nano-structured accelerators within
the fuel, which can be considered structured foams. These
nanostructures are heated using ultra-short, high-contrast laser
pulses, resulting in Coulomb explosions \cite{ruhlkornarXiv1}. The
radiation, fast electrons, and fast ions generated by these Coulomb
exploding nanostructures are then absorbed on picosecond time scales
in the surrounding fuel. The heating concept resembles a combined
electron and ion based fast igniter
\cite{tabak1994ignition,roth2001fast}. This approach offers a new
pathway for achieving controlled fusion reactions with enhanced
efficiency and potential commercial viability. The technology has the
potential to create unique density, velocity, and temperature profiles
within the fuel volume without the need for fuel pre-compression. The
creation of specific density, velocity, and temperature profiles are
known to be a prerequiste for high gain fusion reactor designs
\cite{kidder1968application,duderstadt1982inertial}.

A comprehensive overview of the inertial confinement fusion (ICF)
approach to nuclear fusion is provided in Atzeni and Meyer-ter-Vehn
\cite{atzeni2004physics} and the references therein. The fuel yield,
denoted as $Q_F$, is defined as the ratio of the fusion energy $E_f$
generated to the external energy $E_i$ deposited in the target.
In the literature, as discussed in Abu-Shawareb et
al. \cite{abu2022lawson}, the fusion gain $Q_F$ is distinguished from
the target gain $Q_T = \eta \, Q_F$, where $\eta$ represents the
energy deposition efficiency in the target. The parameter $H$ is
commonly used in the ICF context as a measure of fuel
reactivity, as described in Atzeni and Meyer-ter-Vehn
\cite{atzeni2004physics}. In this paper, $kT_e$ is used to denote the
electron temperature, while $kT_i$ represents the ion
temperature. Typically, $kT_e < kT_i$ due to radiation losses, as
stated in Moreau et al. and Putvinski et
al. \cite{moreau1977potentiality,putvinski2019fusion}.

In order to determine the fusion gain $Q_F$, it is necessary to
calculate the effective burn fraction $\Phi$. The calculation of
$\Phi$ is relatively straightforward when certain simplifying
assumptions are made, such as assuming uniform electron temperature
$kT_e$ and ion temperature $kT_i$, and spatially homogeneous fuel in
either cylindrical or spherical geometry. In the paper, we
focus on cylindrical geometry and enhance the analytical model by
incorporating in-situ fusion energy feedback and implementing a fuel
enclosure. These improvements allow for a more accurate estimation of
$\Phi$ and provide valuable insights into the performance of the
system.

We propose that the energy and power deposition needs of a reactor
operating at $Q_T > 1$ can be fulfilled by utilizing a uniform high
power fuel heating profile without the requirement for fuel
pre-compression. This can be achieved through the synergistic
combination of short-pulse lasers and embedded nano-structured
accelerators featuring small structure sizes. We suggest the use of
materials such as boron composites, which possess the ability to
chemically bind high concentrations of protons, deuterons, and tritium
while maintaining a solid state. By implementing such
nano-accelerators, we can effectively facilitate the desired uniform
energy and power deposition within the reactor.

In cylindrical geometry and neglecting hydro-motion with the exception
of fuel rarefaction, the energy $E_i$, the $\rho_p R$ (density times
radius), and the confinement time $\Delta \tau$ necessary to achieve
the desired fuel gain $Q_F$, as described in \cite{atzeni2004physics},
scale in the following manner
\begin{eqnarray}
  \label{explain1}        
  &&\rho_p R 
  > \frac{3 \, kT_i \, Q_F \, H}{\epsilon_f - 3 \, kT_i \, Q_F} \, , \\
  \label{explain2}  
  &&\Delta \tau 
  > \frac{1}{4 u^s \rho_p} \, \rho_p R \, , \\
  \label{explain3}
  &&\frac{E_i \rho_p}{L}
  >
  \frac{3 \pi \, kT_i}{m_p} \, \left( \rho_p R \right)^2 \, .
\end{eqnarray}
In the given equations, $\rho_p$ represents the mass density of protons
in the fuel used for normalization, $u^s$ denotes the sound velocity,
$\epsilon_f$ represents the fusion energy without neutrons for each
elementary process, $R$ is the radius of the fuel, and $L$ refers to
the length of the fuel cylinder. When the product $kT_i \, Q_F \, H$ is
large, it indicates the requirement for a high ignition energy $E_i$
based on the underlying model described by equations (\ref{explain1})
to (\ref{explain3}).

The energy $E_i$ necessary to achieve the fuel yield $Q_F$ must be
delivered to the fusion plasma within a time duration shorter than the
confinement time $\Delta \tau$. According to (\ref{explain2}) and
(\ref{explain3}) this implies
\begin{eqnarray}
\label{powlow}
  &&P_{\Delta \tau}
  \gg
  \frac{12 \pi \, u^s \, kT_i}{m_p} \, \rho R \, .
\end{eqnarray}
Equation (\ref{powlow}) indicates that a high product of $kT_i \, Q_F \,
H$ necessitates a high deposition power. However, there is an
additional power constraint that arises from the limitations imposed
on the nanostructures. The nanostructures can withstand intense
laser fields for only a brief period of time $\tau$, leading to
\begin{eqnarray}
  &&P_{\tau}
  \approx \frac{E_i}{\tau}
  =
  \frac{kT_i}{m_p \, \tau \, \rho^2_p} \, \left( \rho R \right)^3
  \gg P_{\Delta \tau} \, .
\end{eqnarray}
Therefore, it is necessary to ensure that the power supplied by the
laser system, denoted as $P_L$, exceeds the power constraint
$P_{\tau}$ implying $P_L > P_{\tau}$. Another constraint is
\begin{eqnarray}
\label{powlow1}
  &&P_{d}
  \gg
  \frac{12 \pi \, u^s \, kT_i}{m_p} \, \rho R \, ,
\end{eqnarray}
where $P_d$ is the energy deposition power of the electrons
and ions in the fuel accelerated by the laser irradiated nano-rods and
the deposition power of the associated radiation.

The scaling model presented in (\ref{explain1}) - (\ref{explain3})
provides limited insights into burn requirements. To fully comprehend
burn requirements, it is essential to take into account fusion power
gain and the transfer of power from ions to electrons, which limits
radiation loss. Further details regarding this aspect are elaborated
in section \ref{cond}.

Mixed fuels provide a promising solution to eliminate the need for
cryogenic fuel technology. By employing chemical compounds of boron
that can effectively bind deuterium and tritium at room temperature in
solid form, the challenges associated with cryogenic fuel handling can
be overcome. The fusion performance of the fuel is closely related to the
density, velocity, and temperature profiles established within it 
\cite{kidder1968application}.

\begin{figure}[ht]
\begin{center}
\includegraphics[width=25mm]{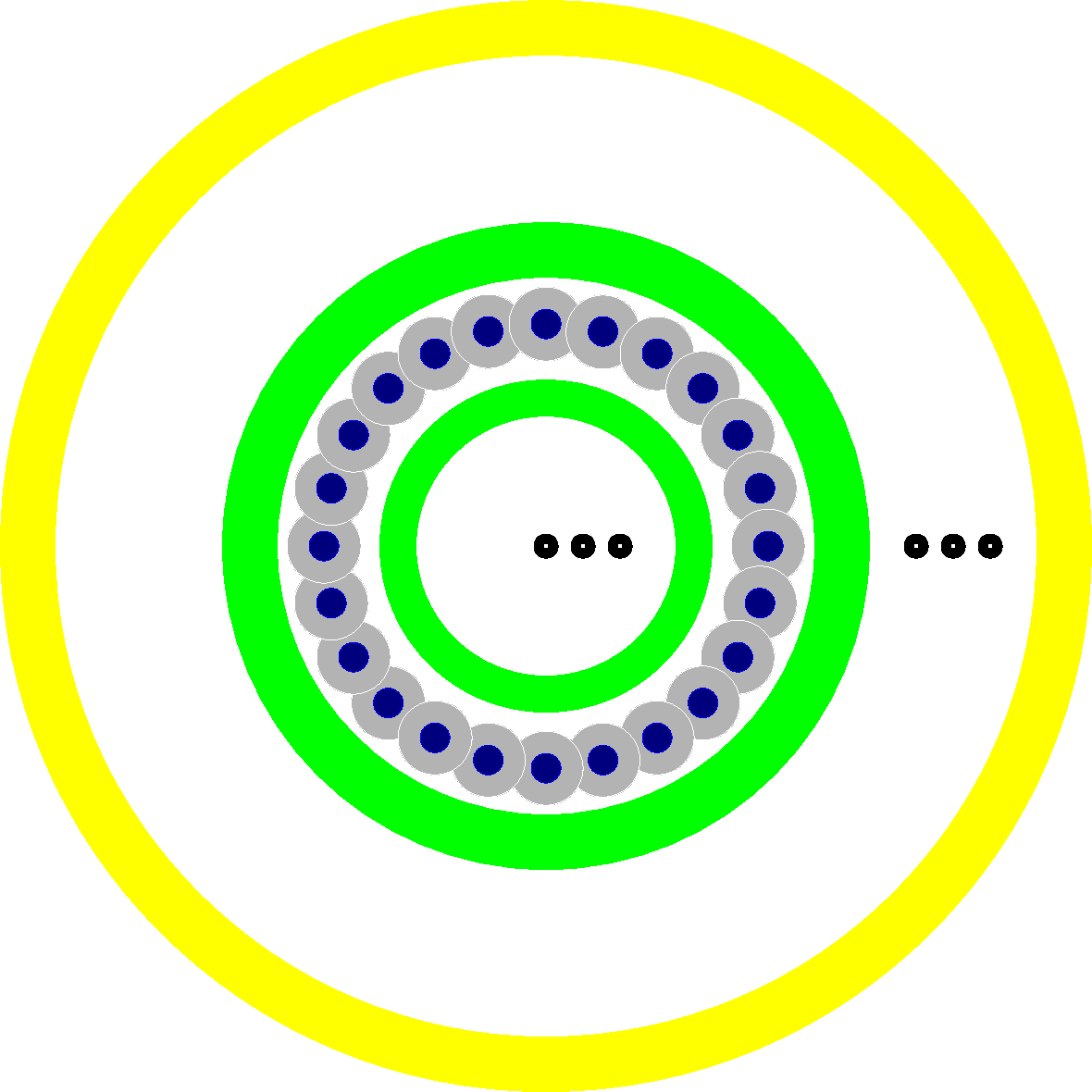}
\hspace{0.5cm}  
\includegraphics[width=10mm]{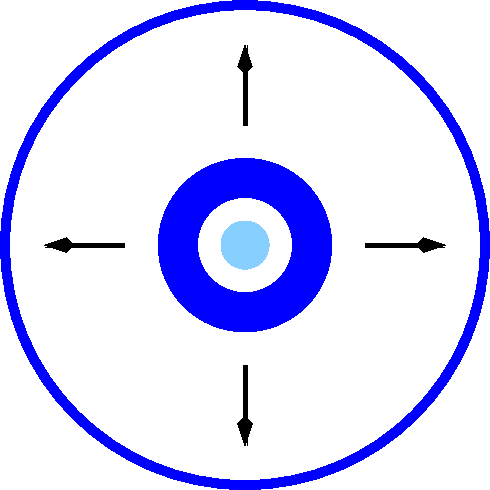}
\hspace{0.5cm}  
\includegraphics[width=25mm]{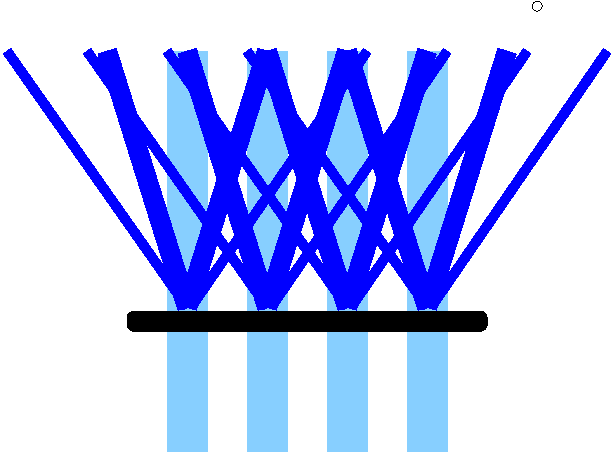} \\
\end{center}
\caption{\label{irrad_rod} The left plot in Figure \ref{irrad_rod}
  showcases the design of a cylindrical onion shell mixed fuel
  reactor, which is enclosed by high-Z materials. This design aims to
  achieve the desired parameters of $\rho_p R$ and confinement time
  $\Delta \tau$. To accomplish this, the reactor incorporates
  acceleration layers consisting of nano-structured $\ce{pBDT}$ fuel
  (represented by blue and grey dots), which are alternated with
  higher density absorbing layers composed of unstructured $\ce{pBDT}$
  fuel (depicted by green lines). Each layer can have a variable
  thickness $\Delta R$. In the design, the laser pulses are incident
  from the top and are depicted as grey disks. The central and right
  plots illustrate the Coulomb explosion of a cluster of embedded
  nano-rods. In the right plot, the short-pulse laser is represented
  by the black bar, while the accelerated fuel ions are depicted by
  the dark blue bars. It is important to note that each rod in the
  cluster has the capability to absorb approximately $10 \, \text{mJ}$
  of energy and deliver a laser deposition power of up to $10.0 \,
  \text{GW}$.}
\end{figure}

In this paper, we propose a method to achieve a uniform temperature
profile in the fuel by merging multiple clusters of nano-rods, which
are interlaced with the fuel, into a single reactor. This reactor is
then surrounded by high-Z materials, as illustrated in Figure
\ref{irrad_rod}. To initiate the fusion process, the fuel is
irradiated with multiple ultra-short, high contrast laser pulses. This
approach enables precise control over the temperature distribution
within the fuel, facilitating fast and efficient energy deposition in
the fuel and enhancing fusion reactions.

Figure \ref{irrad_rod} illustrates a hypothetical cylindrical enclosed
reactor configuration. The design of the reactor includes sections
that are specifically optimized for laser energy deposition, which are
interspersed with fuel absorber sections. The accelerator sections are
composed of nano-structured fuel, serving as embedded accelerators,
while the absorber sections do not possess nano-structures by default.
In the figure, the laser pulses are represented as grey disks,
impinging from the top.

As discussed in \cite{ruhlkornarXiv1}, the use of small nano-rods is
motivated by their ability to deliver the required power density in
the form of quasi-neutral ionic fuel flows, fast electrons, and
radiation. These nano-rods play a crucial role in achieving efficient
energy transfer and deposition within the reactor and to establish
fusion enhancing density, velocity and temperature profiles
\cite{kidder1968application} without fuel pre-compression.

As discussed in \cite{ruhlkornarXiv1}, nano-structures have shown
excellent efficiency in absorbing short-pulse lasers, allowing for
almost complete energy transfer to electrons, ions, and radiation. The
laser-accelerated ions, electrons, and radiation can generate
sufficiently high values of $kT_e$ and $kT_i$.

The deposition power of a single rod exposed to a short laser pulse
can reach up to $10 , \text{GW}$, while the absorbed energy by a
single rod is typically in the range of a few millijoules
($\text{mJ}$). By combining a large number of embedded rods, the
integrated deposition power and absorbed energy can be increased to
meet the desired levels for any given reactor parameters, without
triggering optical instabilities. The energy deposition achieved
through arrays of short laser pulses and small-sized nano-structures
enables the provision of the necessary power densities for
high-efficiency ($\eta$) reactor designs without the requirement for
fuel pre-compression.

The paper is organized as follows. Section \ref{burn} revisits the
burn fraction $\Phi$ and its significance. Section \ref{mixed} focuses
on the investigation of scaling relations for mixed fuel. In section
\ref{cond}, we delve into the discussion of in-situ fusion energy
feedback. Finally, in section \ref{sum}, we provide a summary of our
findings and draw conclusions based on the results obtained.

\section{Revisiting the burn fraction}\label{burn}
As discussed in \cite{ruhlkornarXiv} the elementary burn fraction
$\phi$ in the equal density limit of fuels $n_k$ and $n_l$ on a
microscopic level is
\begin{eqnarray}
  \label{phim}
  \phi_{kl}
  &=& \frac{\Delta n_{k}}{n_k}
  \approx \frac{n_l \, \sigma^{kl}_{R0} \, {\cal R}}{1+ n_l \, \sigma^{kl}_{R0} \, {\cal 
  R}}
\end{eqnarray}
with
\begin{eqnarray}
  \label{sigma-range}
  \sigma^{kl}_{R0} \, {\cal R}
  &=&\int^{\Delta t}_0 d\tau \, u_{kl}(\tau) \, 
   \sigma^{kl}_R  \left( u_{kl}(\tau) \right) \, ,
\end{eqnarray}
where $\sigma_{R0}$ is the effective cross section of the reactive fluids
$k$ and $l$ and $\Delta t$ is the effective resistive stopping time.
The detailed calculation of (\ref{phim}) and (\ref{sigma-range})
requires advanced fluid kinetic simulations as is outlined in
\cite{ruhlkornarXiv1}.

\begin{figure}[ht]
\begin{center}
\includegraphics[width=80mm]{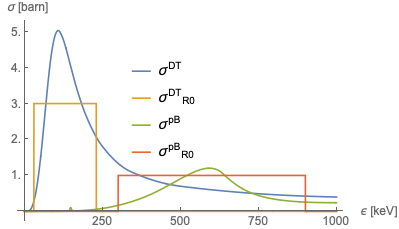}\\
\end{center}
\caption{\label{sigma_approx} The cross sections $\sigma^{DT}$ and
  $\sigma^{pB}$ as functions of energy are illustrated. The yellow and
  purple shading highlights the piecewise constant approximations of
  the cross sections $\sigma^{DT}$ and $\sigma^{pB}$. In the
  present analysis, the approximate cross sections $\sigma^{DT}_{R0}$
  and $\sigma^{pB}_{R0}$ used are the piecewise constant
  approximations.}
\end{figure}

We briefly revisit the ICF paradigm in our study. We propose that by
depositing short-pulse laser energy into the nano-rods of the reactor,
efficient and rapid fuel heating can be achieved, as depicted in
Fig. \ref{irrad_rod}. If a sufficiently long confinement time $\Delta
\tau$ can be established, the ions can be heated to the critical
temperature $kT_i$, while the electrons can reach the temperature
$kT_e$ in a fraction of the latter time.

Under these assumptions, considering Maxwellian distributions of
electrons and ions with uniform temperatures $kT_e$ and $kT_i$, and
neglecting hydro-motion, we can calculate the reactivity
\begin{eqnarray}
  \label{cross_velo}
 \sigma^{kl}_{R0} \, R
 &\approx& \Delta \tau \, u_{kl}\, \sigma^{kl}_{R0} \, , 
\end{eqnarray}
where the resistive range ${\cal R}$ has been replaced by the fuel
radius $R$ in the assumed cylindrical geometry. The emergence of the
radius $R$ occurs when the reactor reaches equilibrium. The parameter
$R \gg {\cal R}$ is associated with the confinement time $\Delta
\tau$, which is primarily limited by fluid rarefaction at leading
order, approximated by
\begin{eqnarray}
\label{conf_time}
  \Delta \tau &\approx& \frac{R}{4 \, u^s} \, , \quad
  u^s > \sqrt{\frac{3 \, kT_Z}{m_Z}} \, .
\end{eqnarray}
Here, $m_Z$ is an effective mass intended to account for the effective
inertia of the fuel enclosure and $kT_z$ is the effective temperature of the
latter. The parameter $u^s$ is an effective sound velocity based on
the enclosure temperature $kT_Z$ and $m_Z$. The burn fraction $\Phi$
in the context is
\begin{eqnarray}
  \label{phi}
  \Phi_{kl}
  &\approx&
   \frac{\rho_p R}{H_{kl} + \rho_p R}
\end{eqnarray}
with
\begin{eqnarray}
\label{HH}
H_{kl} \left( kT_i, u^s \right)
&\approx&
\frac{4 \, m_p \, u^s}{\sigma^{kl}_{R0} \, u_{kl}}  \, ,  
\end{eqnarray}
where $H$ is normalized to the proton mass $m_p$ without loss of
generality. The $\sigma u$ required in (\ref{HH}) is approximated as
follows
\begin{eqnarray}
  \label{reactivity}
  &&\left( u_{kl} \, \sigma^{kl}_R \right) \left( kT_i \right) \\
  &\approx&\sqrt{\frac{8 \, kT_i}{\pi \, m_{kl}}} \, \sum^N_{i=1}
            \sigma^{kl}_{2i-1} \, \left[ \left(
            1+\frac{\epsilon^{kl}_{2i-1}}{kT_i}
            \right) \, e^{-\frac{\epsilon^{kl}_{2i-1}}{kT_i}}
            \right. \nonumber \\
  &&\hspace{3.5cm} \left. - \left( 1+ \frac{\epsilon^{kl}_{2i}}{kT_i}
      \right) \, e^{-\frac{\epsilon^{kl}_{2i}}{kT_i}} \right] \, , \nonumber
\end{eqnarray}
where
\begin{eqnarray}
  \label{reduced_mass}  
  m_{kl}&=&\frac{m_k \, m_l}{m_k + m_l} \, , \\
  \label{cross_sec_R0}
  \sigma^{kl}_R \left( \epsilon \right) &\ge& 
   \left\{
   \begin{array}{ll}
     \sigma^{kl}_{2i-1} \, , & \epsilon^{kl}_{2i-1} \le 
     \epsilon \le \epsilon^{kl}_{2i} \\
     0 \, , & \text{else} \\
   \end{array}
  \right. 
\end{eqnarray}
for $i=1,2,3, ...$ as is illustrated by the piecewise constant 
functions $\sigma^{DT}_{2i-1}$ and $\sigma^{pB}_{2i-1}$ in 
Fig. \ref{sigma_approx} for the fuels $\ce{DT}$ and 
$\ce{pB}$. The parameters $m_k$ and $m_l$ are the masses of
particles $k,l=\ce{pBDT}$.

It is important to highlight that the replacement of ${\cal R}$ with
$R$ in the hydrodynamic context is a result of thermal
quasi-equilibrium. This thermal equilibrium is necessary because
Coulomb collisions occur much more frequently than fusion
collisions. While Coulomb collisions are still abundant in thermal
equilibrium, they do not significantly alter hydrodynamic parameters
on hydro timescale as soon as the latter has been reached. By
establishing long confinement times for the heated fuel by
technological means, which is typically in a rare fusion state, there is a
possibility for fusion events to become more abundant, potentially
leading to sufficient fusion gain. On the other hand, non-equilibrium
fuels tend to dissipate their energy into collisional degrees of
freedom, which suppresses fusion yield. In a broader context, this
implies that high gain beam fusion is not feasible but may be useful
for fusion reactor designs that are not intended for energy production.

\section{Lower threshold for the density - range product}\label{mixed}
The understanding of the potential of mixed fuels to generate fusion
energy is still incomplete. Additionally, the proposed direct drive
fast heating technology opens up the possibility of achieving
uniformly heated fuel volumes without fuel pre-compression. More
advanced initial density, velocity, and temperature profiles may play
a crucial role in achieving high gain targets, as highlighted in
Kidder et al. \cite{kidder1968application}.

Therefore, it is crucial to establish lower thresholds for the
density-range products of potential fuel mixtures for the fuel gains
$Q_F$ at a given temperatures $kT_e$ and $kT_i$. The density-range
product depends on various parameters, including $\epsilon_f$, $Q_f$,
$H$, and $kT_i$, as expressed in equations (\ref{explain1}) -
(\ref{explain3}). In the context of mixed fuels, these equations need
to be generalized.

\begin{figure}[ht]
\begin{center}
  \includegraphics[width=70mm]{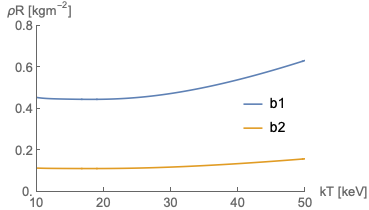}\\
\end{center}
\caption{\label{rhoRDTpBmix} $\rho_p  R$ for $Q_F = 1.0$
  with $b_1$: $n_p=n_D=n_T= n_B= 1.5 \cdot 10^{29} \,
  \text{m}^{-3}$, $u^s=\sqrt{3 \, kT_i/m_B}$, and
  $b_2$: $n_p=n_D=n_T= n_B=1.5 \cdot 10^{29} \,
  \text{m}^{-3}$, $u^s=\sqrt{3 \, kT_i/16 \, m_B}$.}
\end{figure}

\begin{figure}[ht]
\begin{center}
\includegraphics[width=70mm]{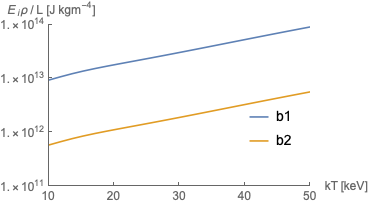}\\  
\end{center}
\caption{\label{EiDTpBmix} $E_i \rho_p/L$ for $Q_F = 1.0$
  with $b_1$: $n_p= n_D= n_T= n_B=1.5 \cdot 10^{29} \,
  \text{m}^{-3}$, $u^s=\sqrt{3 \, kT_i/m_B}$, and
  $b_2$: $n_p=n_D=n_T=n_B=1.5 \cdot 10^{29} \,
  \text{m}^{-3}$, $u^s=\sqrt{3 \, kT_i/16 \, m_B}$.}
\end{figure}

Let's consider a fuel mixture of $\ce{pBDT}$ with the number densities
$n_p$, $n_B$, $n_D$, and $n_T$, where $n_p$ represents the proton
density, $n_B$ represents the boron density, $n_D$ represents the
deuterium density, and $n_T$ represents the tritium density. Utilizing
the ideal gas equation of state and the reactive rate equations, we
obtain
\begin{eqnarray}
  \frac{dn_p}{dt}
  &\approx& - n_p \, n_B \, u_{pB} \,  \sigma^{pB}_R \, , \\
  \frac{dn_D}{dt}
  &\approx& - n_D \, n_T \, u_{DT} \,  \sigma^{DT}_R \, ,
\end{eqnarray}
where the thermal reactivities are given by (\ref{reactivity})
and
\begin{eqnarray}
 \label{EiDTpB}
 E_{i}
  &\approx&
  \frac{3\pi L \, kT_i}{2 \rho^2_p} \, \left( n_p + n_B + n_D + n_T  \right) 
            \, \left( \rho_p R \right)^2 \, , \\
  E_f
  &\approx&
  \frac{\pi L \, \epsilon^{DT}_f}{m_p \, \rho_p} \,
  \frac{\left( \rho_p R \right)^3}{H_{DT} \left( kT_i,
            u^s \right) + \rho_p R} \\
  &&
  +\frac{\pi L \, \epsilon^{pB}_f}{m_p \, \rho_p} \,
  \frac{\left( \rho_p R \right)^3}{H_{pB} \left( kT_i,
     u^s \right) + \rho_p \, R} \, , \nonumber
\end{eqnarray}
where $E_i$ is the initial energy in the reactor and the fusion 
energy is $E_f$. The parameters $H_{DT}$ and $H_{pB}$ can be obtained
from (\ref{HH}). The fuel yield $Q_F$ is
\begin{eqnarray}
  Q_F
  &=&
   A_1\, \frac{ \rho_p R}{H_{DT} \left( kT_i,
      u^s \right) + \rho_p R} \\
  &&+A_2 \, \frac{ \rho_p R}{H_{pB} \left( kT_i,
     u^s \right) + \rho_p R} \, , \nonumber
\end{eqnarray}
where
\begin{eqnarray}
A_1
&=&
\frac{2 \, \epsilon^{DT}_f \, n_D}{3 \, kT_i \, \left( n_p + n_D + n_T +
    n_B \right)} \, \\
A_2
&=&
\frac{2 \, \epsilon^{pB}_f \, n_p}{3 \, kT_i \, \left( n_p+ n_D + n_T
    + n_B \right)} \, .
\end{eqnarray}
We introduce  
\begin{eqnarray}
&&H_1 = H_{DT} \left( kT_i,
     u^s \right)  \, , \quad H_2 = H_{pB} \left( kT_i,
     u^s \right)
\end{eqnarray}   
and obtain for the lower limit of $\rho_p R$ required to reach the
fuel yield $Q_F$ at the temperatures $kT_e$ and $kT_i$
\begin{eqnarray}
  \label{rhoRDTpB2}
  &&\hspace{-1cm}\rho_p R \\
  &&\hspace{-1cm}\ge
  \frac{\left( H_1 + H_2 \right) \, Q_F - \left( A_1 H_2 + A_2 H_1
      \right) }{2 \, \left( A_1 + A_2 - Q_F \right)} \nonumber \\
  &&\hspace{-1cm}+ \sqrt{\frac{H_1 H_2 Q_F}{A_1+A_2 -Q_F} +\left(  \frac{\left( H_1 +
     H_2 \right) \, Q_F - \left( A_1 H_2 + A_2 H_1 \right)}{2 \, \left( A_1 + A_2 -
     Q_F \right)} \right)^2} \nonumber
\end{eqnarray}
where
\begin{eqnarray}
Q_F&\le& A_1 + A_2 \, .
\end{eqnarray}
The scaled ignition energy $E_i \rho_p/L$ is obtained by plugging
(\ref{rhoRDTpB2}) into (\ref{EiDTpB}). The scalings of $E_i
\rho_p/L$ and $\rho_p R$ for $Q_F = 1.0$ with reactor temperature
$kT_i$ according (\ref{EiDTpB}) and (\ref{rhoRDTpB2})
are illustrated in Figs. \ref{rhoRDTpBmix} and \ref{EiDTpBmix} for
two different effective sound velocities obtained for $m_Z=m_B$ and
$m_Z=16 \, m_B$, where $m_B$ is the boron mass.

Other energies $E_i$ required for fuel yields $Q_F \ne 1$ as functions  
of $kT_i$ can be obtained from the $Q_F = 1.0$ case by rescaling the  
curves in Figs. \ref{rhoRDTpBmix} and \ref{EiDTpBmix} with the help of  
(\ref{EiDTpB}) and (\ref{rhoRDTpB2}) and adapting the number density
fractions as required.  

It is evident from Figs. \ref{rhoRDTpBmix} and \ref{EiDTpBmix} that
mixed fuel reactor concepts neglecting in-situ fusion energy feedback
and measures improving confinement require unsustainably large $E_i$
for $Q_F \gg 1$ without fuel compression. Since we are not interested
in high gain concepts in the present paper, it is obvious that
confinement has to be improved implying larger effective mass $m_z$
and fusion energy has to be fed back into the fuel.

We analyse aspects of in-situ fusion energy feedback
in section \ref{cond}.

\section{Aspects of in-situ fusion energy feedback}\label{cond}
In this section, we investigate the influence of in-situ fusion energy
feedback on the scaled parameter $E_i , \rho_p/L$. Our analysis
focuses on determining the necessary electron temperature $kT_e$ and
ion temperature $kT_i$ for achieving burn through a simple numerical
model. It is important to note that we plan to enhance this model in
future papers.

We first analyze the frequency integrated normalized electronic
radiation power density in a plasma, which is approximated by
\cite{befki1966radiation}
\begin{eqnarray}
  \label{radloss}
  \frac{P_r}{\rho^2_p}
  &\approx&
  \frac{16}{3 \hbar \, m^2_p} \, \left( \frac{e^2}{4\pi \epsilon_0} \right)^3
  \, \frac{\sqrt{kT_e} \, G}{n^2_p \, \left(m_ec^2
            \right)^{\frac{3}{2}}} \\
  &&\times \, \sum_{l=D,T,p,B,\alpha} Z^2_l n_l n_e \, , \nonumber
\end{eqnarray}
where $G > 1$ is assumed and can be adapted if necessary. Next, we
need the effective normalized fusion power density $P_f$, which is
approximated as
\begin{eqnarray}
  \label{Pfus}
  \frac{P_f}{\rho^2_p}
  &\approx&
  a \, \epsilon^{DT}_f \, \frac{n_D \, n_T}{m^2_p \, n^2_p} \, u_{DT}
            \, \sigma^{DT}_{R0} \\
  &&+ a\, \epsilon^{pB}_f \, \frac{n_p \, n_B}{m^2_p \, n^2_p} \,
     u_{pB} \, \sigma^{pB}_{R0} \, , \nonumber
\end{eqnarray}
where the $a \, \epsilon$ in (\ref{Pfus}) with $0 \le a < 1$ only
account for the fusion energy deposited in the fuel. This implies that
the energies of the fusion neutrons are neglected for the deposition of
fusion energy in the fuel. The averaged reactivity $\sigma_{R0} u$ is
given by (\ref{reactivity}). One obvious requirement for burn is
\begin{eqnarray}
  \label{tmax}
  \frac{P_f}{\rho^2_p} &\ge&\frac{P_r}{\rho^2_p}
\end{eqnarray}
implying an upper limit for the electron temperature $kT_e$ according
to (\ref{radloss}), at which $P_r/\rho^2_p$ exceeds $P_f/\rho^2_p$.

We make the assumption that the transfer of energy from ions to
electrons is the sole mechanism for electron energy gain in the
reactor. Therefore, it is necessary to determine the rate of energy
transfer from ions to electrons. Based on the work of Moreau and
Spitzer \cite{moreau1977potentiality, spitzer2006physics}, the power
transfer normalized by density from ions to electrons in the case of
mixed fuels can be approximated as follows
\begin{eqnarray}
  \label{deposi}
  \frac{P_{ie}}{\rho^2_p}
  &\approx&\sum_{l=D,T,p,B,\alpha} \frac{n_l}{m^2_p \, n^2_p \, t^{eq}_{le}} \, \left(
  kT_i -kT_e \right) \, ,
\end{eqnarray}
where
\begin{eqnarray}
  t^{eq}_{le}
  &\approx& \frac{4 \pi \epsilon^2_0 \, m_l \, m_e}{Z^2_l \, q^4_e \, n_e \, \ln \Lambda
         \left( n_l, n_e \right)} \, \left( \frac{kT_i}{m_l} +
  \frac{kT_e}{m_e} \right)^{\frac{3}{2}}
\end{eqnarray}
are the Spitzer equilibration times, where is approximation of
Maxwellian distributions for all particles is made.

To analyze the Spitzer theory of plasma heating by
$\alpha$-particles in more detail we introduce the temperature
$kT_{\alpha}$ to distinguish the temperature of the $\alpha$-particles
from the general background, which is assumed to have the temperature
$kT$. Making the simplifying assumption $kT_e = kT_i = kT$ we obtain
with the help of (\ref{deposi}) for the ratios of the energy transfer
rates between $\alpha$-particles and arbitrary ions $i$ and
$\alpha$-particles and electrons
\begin{eqnarray}
  \frac{P_{\alpha i}}{P_{\alpha e}}
  &=&\frac{n_e \, m_e}{n_i \, m_i}
  \, \left(\frac{ 1+
      \frac{m_{\alpha} \, kT}{m_e \, kT_{\alpha}}}{1 +\frac{m_{\alpha} \,
      kT}{m_i \, kT_{\alpha}}} \right)^{\frac{3}{2}} \, .
\end{eqnarray}
There are a few cases that can be distinguished easily
\begin{eqnarray}
  \label{tgapa}
  \frac{m_i}{m_{\alpha}} < \frac{kT}{kT_{\alpha}}
  &\rightarrow&  \frac{P_{\alpha i}}{P_{\alpha e}}
  \approx \frac{n_e}{ n_i} \, \sqrt{\frac{m_i}{m_e}} \gg 1 \, , \\
  \label{tgapm}
  \frac{m_i}{m_{\alpha}} > \frac{kT}{kT_{\alpha}} > \frac{m_e}{m_{\alpha}}
  &\rightarrow&
  \frac{P_{\alpha i}}{P_{\alpha e}}
  \approx \frac{n_e \, m^{\frac{3}{2}}_{\alpha}}
  { n_i \, m_i \, \sqrt{m_e}} \, \left( \frac{kT}{kT_{\alpha}}
  \right)^{\frac{3}{2}} \, , \\    
  \frac{m_e}{m_{\alpha}} > \frac{kT}{kT_{\alpha}}
  \label{tgape}
  &\rightarrow&  \frac{P_{\alpha i}}{P_{\alpha e}}
  \approx \frac{n_e \, m_e}{ n_i \, m_i} \ll 1 \, .
\end{eqnarray}
For $kT \gg kT_{\alpha}$, see (\ref{tgape}), the power transfer from
$\alpha$-particles into ions is much larger than the power transfer
form $\alpha$-particles into electrons. For $kT \ll kT_{\alpha}$
predominantly the electrons are heated by the $\alpha$-particles.
There is, however, a limit to electron heating by $\alpha$-particles
or ions in general. According to (\ref{deposi}) ions cannot heat electrons 
as soon as $kT_e > kT_i$ holds.

To gain a comprehensive understanding of the limitations associated
with the energy deposition caused by $\alpha$ particles and other ions
in a fusion plasma, it is insufficient to focus solely on the relative heating
power transferred to electrons and ions in the background plasma at an
ambient temperature of $kT$. It is crucial to estimate the range of
these particles and their total energy deposition capacity within the
plasma. For information on $\alpha$-particle stopping power models for
$\ce{DT}$, refer to the discussion found in \cite{xu2014effects}. A more
general discussion of the stopping power excerted on ions in partially
ionized matter and hot plasma is found in \cite{basko1983stopping}.
From \cite{basko1983stopping}, we quote the stopping power exerted by
free electrons with the number density $n_e$ at the temperature $kT_e$
on a single ion and obtain the following equations of motion
\begin{eqnarray}
  \frac{dr_i}{dt} &=& v_i \, , \\
  \frac{d v_i}{dt} &\approx&-S_{fe} \, ,
\end{eqnarray}
where $r_i$ is the ion position and $v_i$ its velocity and
\begin{eqnarray}
S_{fe} &\approx& \frac{Z^2_i e^4 \, n_e \, G \left( x_e
          \right)}{4 \pi \epsilon^2_0 \, m_{ei} m_i \, v^2_i} \, \ln \Lambda_{fe} \, , \\
  G \left( x_e \right) &\approx&\frac{2}{\sqrt{\pi}} \, \left( \int^{x_e}_0
                           dt \, e^{-t^2} - x_e \, e^{-x^2_e} \right)
                           \, , \\
  x_e &\approx&\sqrt{\frac{m_e v^2_i}{2 \, kT_e}} \, , \\
  m_{ei}&=&\frac{m_em_i}{m_e+m_i} \, .
\end{eqnarray}
The parameter $m_i$ is the ion mass and $Z_i$ the ion charge
number. The electron density $n_e$ depends on the ionic background.

\begin{figure}[ht]
\begin{center}
  \includegraphics[width=0.8\linewidth]{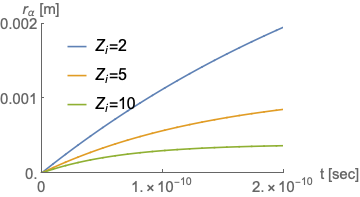}\\
\end{center}
\caption{\label{stopping_range} Ranges of various ions including $\alpha$-particles 
 as functions of time with initial energy $\epsilon_{i} = 3.5 \, \text{MeV}$ in the 
 $\ce{pBDT}$ fuel mix at $kT_e = 20 \, \text{keV}$ with the number 
 densities $0.6 \, n_p =  0.6 \, n_D = 0.6 \, n_T = n_B$, where 
 $\rho_p \approx 300 \, \text{kgm}^{-3}$.}
\end{figure}

\begin{figure}[ht]
\begin{center}
  \includegraphics[width=0.8\linewidth]{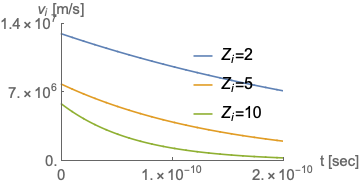}\\  
\end{center}
\caption{\label{stopping_velocity} Velocities of various ions including $\alpha$-particles 
 as functions of time with initial energy $\epsilon_{i} = 3.5 \, \text{MeV}$ in the 
 $\ce{pBDT}$ fuel mix at $kT_e = 20 \, \text{keV}$ with the number 
 densities $0.6 \, n_p =  0.6 \, n_D = 0.6 \, n_T = n_B$, where 
 $\rho_p \approx 300 \, \text{kgm}^{-3}$.}
\end{figure}

Figures \ref{stopping_range} and \ref{stopping_velocity} illustrate
the range and velocity of various ions with an initial
energies of $\epsilon_{i} =3.5 , \text{MeV}$ in a $\ce{pBDT}$ mixed
fuel background at $kT_e = 20 , \text{keV}$. It is evident that for
fuel radii ranging from $R = 1 - 2 \, \text{mm}$, significant
fractions of the energies of the ions from the embedded accelerators
and of the $\alpha$ particles can be deposited in the
plasma. The range of the $\alpha$ particles implies that
the scaled parameter $E_i , \rho_p/L$ will significantly decrease by
incorporating in-situ fusion energy feedback, as compared to
Fig. \ref{EiDTpBmix}.

We determine lower temperature thresholds, specifically $kT_e < kT_i$,
at which the deposited normalized fusion power surpasses the
normalized power transfer from ions to electrons by equating 
\begin{eqnarray}
\label{tgap}
&&\frac{P_f}{\rho^2_p} = \frac{P_{ie}}{\rho^2_p} =
   \frac{P_{r}}{\rho^2_p} \, .
\end{eqnarray}
It is important to note that this temperature threshold also serves as
the upper limit for the normalized radiation power.

\begin{figure}[ht]
\begin{center}
\includegraphics[width=80mm]{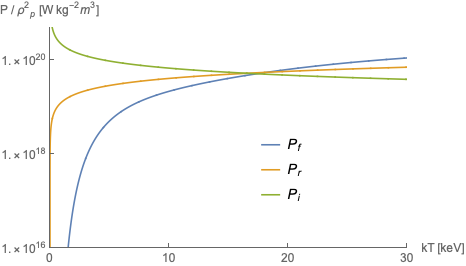}\\  
\end{center}
\caption{\label{PDTpBmixfus} Scaled fusion, radiation, and transfer 
powers for the mixed fuel consisting of $\ce{pBDT}$. It holds 
$P_{f}/\rho^2_p = P_r/\rho^2_p = P_{ie}/\rho^2_p$ at $kT_i 
\approx 18 \, \text{keV}$ and $kT_e \approx 0.73 \, kT_i$ for 
$G =1$. The number densities are $0.6 \, n_p= 0.6 \, n_D = 0.6 \, n_T 
= n_B= 0.5 \cdot 10^{29} \, \text{m}^{-3}$ and $n_e= \sum_{k=p,B,D,T} Z_k n_k$. }
\end{figure}

Figure \ref{PDTpBmixfus} illustrates that for $0.6 \, n_p = 0.6 \, n_D =
0.6 \, n_T = n_B = 0.5 \cdot 10^{29} \, \text{m}^{-3}$, equations
(\ref{tgap}) are satisfied if $kT_e \approx 0.73 \, kT_i$ and $kT_i
\approx 18 \, \text{keV}$, assuming that the total energy of the fusion
$\alpha$-particles is deposited in the fuel and radiation losses are
described by (\ref{radloss}) with $G=1$.

Therefore, if the embedded accelerator technology is capable of
heating fuel ions to $kT_i > 18 \, \text{keV}$ for the fuel number
densities in $\ce{pBDT}$ within a fraction of the confinement time,
implying $\Delta \tau \gg t^{eq}_{ij}$, the ion temperature $kT_i$ can
continue to increase up to an upper limit unless there are additional
loss processes not considered in this section.

\section{Fusion yield including in-situ energy feedback}
In this section, we analyze the influence of in-situ fusion energy
feedback and enhanced inertia on the fusion yield of the mixed fuel
$\ce{pBDT}$. To account for the highly nonlinear impact of in-situ
fusion energy feedback on the ambient fusion plasma, we employ a
simple numerical model that allows radiation to freely exit the
reactor. This approach facilitates the investigation of the effects of
feedback and enhanced inertia on the fusion yield.

By adjusting the available fusion power in the plasma $a P_f$, we can
account for the incomplete deposition of the energy carried by the
$\alpha$-particles in the fuel. To model the influence of a high-$Z$
enclosure, as illustrated in Fig \ref{irrad_rod}, we introduce the
effective temperature $kT_Z = c \, kT_i$ with $0 \le c < 1$ and the
effective fuel mass $m_Z$. These parameters modify fluid rarefaction
in the system. Meanwhile, the fuel temperatures are given by $kT_e = b
, kT_i$, where $0 < b \le 1$, as derived from equations (\ref{tgapa})
- (\ref{tgap}).

Assuming known densities $n_p$, $n_D$, $n_T$, and $n_B$, and the
applicability of ideal equations of state, we consider an initial
energy in the fuel of $E^0_f = E_i$ and an initial electron
temperature of $kT^0_e$. From these conditions, we derive the initial
ion temperature $kT^0_i$ as follows
\begin{eqnarray}
 kT^0_i 
 &=&
 \frac{2 \, \left( E^0_i -\frac{3}{2} \pi LR^2 n_e kT^0_e \right)}{3 \,
 \pi \, LR^2 \, \left( n_p+ n_D + n_T + n_B \right)} \, .
\end{eqnarray}
Next, we make the assumption that the electron temperature $kT_e$
rapidly decreases, resulting in $kT_e = b , kT_i$ with $0 \leq b <
1$. Our reactor configuration is enclosed within high-Z materials, as
illustrated in Figure \ref{irrad_rod}. To approximate hydrodynamic
effects, we employ an effective fluid rarefaction model using the
effective temperature $kT_Z$ and the effective mass $m_Z$. In this
model, we neglect fuel depletion due to reactions. With these
considerations, we utilize the simplified model below to estimate
the fusion yield $Q_F$ incorporating in-situ fusion
energy feedback into our analysis.

We set $\Delta t = 10^{-12} \,
\text{s}$, $m_Z=2.7 \cdot 10^{-25} \, \text{kg}$, $a=0.3$, $b=0.73$,
$c=0.1$, $\rho_{DT}= 500 \, \text{kgm}^{-3}$, $\rho_{pB} = 800 \,
\text{kgm}^{-3}$, $0.6 \, n_p = 0.6 \, n_D = 0.6 \, n_T = n_B$, $kT_e=
20 \, \text{keV}$, and $R=L=1.0 \, \text{mm}$ and iterate
\begin{eqnarray}
  \label{simpledyn}
  t^n
  &=&
  t^n+\Delta t \, , \\
  kT^n_i 
  &=&
  \frac{2 \, \left( E^n_d -\frac{3}{2} \pi LR^2 n_e kT_e \right)}{3 \,
      \pi \, LR^2 \, \left( n_p+ n_D + n_T + n_B \right)} \, , \nonumber \\
  kT^n_e
  &=&
  b \, kT^n_i \nonumber \\
  kT^n_Z
  &=&
  c \, kT^n_i \, , \nonumber \\
  u^n
  &=&
  \sqrt{\frac{3 \, kT^n_Z}{m_Z}} \, , \nonumber \\
  \Delta \tau^n
  &=&
  \frac{R}{4 \, u^n} \, , \nonumber \\
  E^n_d
  &=&
  E^n_d+\pi \, L R^2 \, \Delta t \, \left[ a \, P_f \left( kT^n_i \right) - P_{ie} \left(
      kT^n_i, kT^n_e \right) \right] \, , \nonumber
\end{eqnarray}
where $\Delta \tau^n \gg t^{eq, n}_{ij} $ is assumed and
\begin{eqnarray}
  E^n_f &=& E^n_f + \pi \, L R^2 \, \Delta t \, P_f \left( kT^n_i 
            \right) \, , \nonumber \\
  E^n_{\alpha} &=& E^n_{\alpha} + \pi \, L R^2 \, \Delta t \, P_f \left( kT^n_i 
            \right) \, , \nonumber \\
  E^n_n &=& E^n_n + \frac{14.1}{3.5} \, \pi \, L R^2 \, \left( 1 -a \right) \, \Delta t \, P_f \left( kT^n_i 
            \right) \, , \nonumber \\  
   E^n_r &=& E^n_r + \pi \, L R^2 \, \Delta t \, P_{ie} \left( kT^n_i 
            \right) \, , \nonumber \\
  Q^n_F
  &=&
  \frac{E^n_f}{E_i} \, . \nonumber
\end{eqnarray}
The term $Q_F^n$ represents the fuel yield without neutrons at time
$t^n$. The parameters $E_d$, $E_f$, $E_n$, $E_r$, and $E_{\alpha}$
correspond to the integrated fusion energy deposition in the fuel
ions, the integrated fusion energy, the integrated neutron energy,
the integrated emitted radiation energy, and the emitted
$\alpha$-particle energy at time $t^n$, respectively. The parameter
$a$ denotes the fraction of the energy of the $\alpha$-particles that
is deposited in the fuel. This value is determined using the stopping
power of the $\alpha$-particles in a plasma with temperatures $kT_e^n
= b , kT_i^n$. The parameter $b$ characterizes the degree of electron
heating in the fuel, while the parameters $c$ and $m_Z$ control the
effective confinement of the latter.

The fusion power $P_f$ and the power transfer from ions to electrons
$P_{ie}$ are given by equations (\ref{Pfus}) and (\ref{deposi}),
respectively. The conditions $\Delta \tau^n \gg t^{eq, n}_{ij}$
indicate that the energy equilibration time between ions and electrons
and between ions within the fuel is much shorter than the confinement
time $\Delta \tau^n$. The loop in equation (\ref{simpledyn}) is
terminated when $t^n > \Delta \tau^n$.

\begin{figure}[ht]
  \begin{center}
\includegraphics[width=70mm]{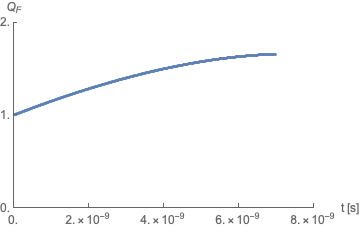}\\  
\end{center}
\caption{\label{fuel_yield} The fuel yield $Q^n_f$ is a time-dependent
  quantity that neglects the contribution of neutrons. The parameters
  used in this context are $0.6 \, n_p  = 0.6 \, n_D = 0.6 \, n_T =
  n_B$, where $\rho_{DT} = 500 \, \text{kgm}^{-3}$ and $\rho_{pB} = 800 \,
  \text{kgm}^{-3}$. The initial temperatures are set to $kT^0_i = 20 \,
  \text{keV}$ and $kT^0_e = 0.73 \, kT^0_i$.}
\end{figure}

\begin{figure}[ht]
\begin{center}
\includegraphics[width=80mm]{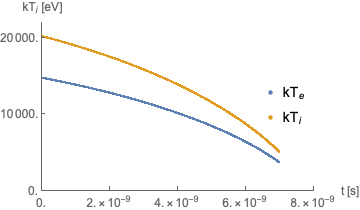}\\
\end{center}
\caption{\label{temperatures} Temperatures $kT^n_e$ and $kT^n_i$ as
  functions of time. The parameters are $0.6 \, n_p
  = 0.6 \, n_D = 0.6 \, n_T = n_B$, where $\rho_{DT} = 500 \,
  \text{kgm}^{-3}$ and $\rho_{pB} = 800 \, \text{kgm}^{-3}$. The
  initial temperatures are $kT^0_i = 20 \, \text{keV}$ and $kT^0_e = 0.73
  \, kT^0_i$.}
\end{figure}

\begin{figure}[ht]
\begin{center}
\includegraphics[width=80mm]{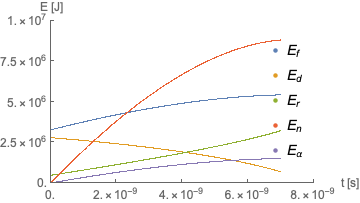}\\  
\end{center}
\caption{\label{energies} Fusion energies as functions of time, where $E^n_f$,
  $E^n_d$, $E^n_r$, $E^n_n$, and $E^n_{\alpha}$ are the integrated fusion, ion
  deposition, radiation loss, neutron loss, and exiting $\alpha$-particle
  energies as functions of time. The other parameters are $0.6 \, n_p
  = 0.6 \, n_D = 0.6 \, n_T = n_B$, where $\rho_{DT} = 500 \,
  \text{kgm}^{-3}$ and $\rho_{pB} = 800 \, \text{kgm}^{-3}$. The
  initial temperatures are $kT^0_i = 20 \, \text{keV}$ and $kT^0_e = 0.73
  \, kT^0_i$.}
\end{figure}

Figures \ref{fuel_yield}, \ref{temperatures}, and
\ref{energies} show the fuel yield, the fuel temperatures $kT_e$ and
$kT_i$, and various ion energies and the neutron energy as functions of
time in the reactor. The neutrons carry the bulk of the generated
fusion energy in the reactor.

\section{Fuel heating by structured foams}
The arrangement of the direct drive fast fuel heating concept for the
proposed low-gain fusion reactor can be modeled using the effective
rod concept, as detailed in the work by Ruhl et
al. \cite{ruhlkornarXiv2}. According to the discussions presented in
this reference, the effective rod model can be envisioned as a
structured foam, suitable for operation under the conditions of a
single effective rod limit. Structured foams exhibit superior laser
energy and power absorption capabilities compared to the unstructured
foams, the homogeneous plasma or the empty hohlraums.

In this context, an effective single rod demonstrates the ability to
efficiently absorb laser energy at a consistent rate per surface area
$A_r$ and per unit length of laser propagation. As elucidated in Ruhl et
al. \cite{ruhlkornarXiv2}, the extent of laser energy and power
absorption is contingent upon factors such as the radius of the rod,
the material it is composed of, and the spacing $\sqrt{A_r}$ between
neighboring rods called the pitch.

In simulations conducted with a specific configuration featuring a rod
diameter of $30 \, \text{nm}$ and a pitch $\sqrt{A_r} \approx 500 \,
\text{nm}$, as outlined in \cite{ruhlkornarXiv2}, the calculated
absorption power density of a single effective rod is approximately
\begin{eqnarray}
\epsilon_r&\approx& 1.2 \cdot 10^{-2} \, \frac{\text{J}}{\mu \text{m}^3} \, . 
\end{eqnarray}
We note that structured foams can be placed and oriented in any
desired way in a given volume.

For the sake of simplicity, we adopt a cylindrical geometry in our
considerations. We make the assumption that the nano-rods are
arranged radially, ensuring even distribution across the area defined
by the radius $R_r$ and the radial with $\Delta R_r$. Additionally,
these rods are assumed to be aligned along the cylinder axis and
possess a uniform length denoted as $\Delta L_r$.

Under the assumptions made the total energy, denoted as $E$, which the
structured foam comprised of these rods can potentially absorb, is
approximately determined by $\epsilon_r$, $R_r$, $\Delta R_r$, and
$\Delta L_r$. We obtain
\begin{eqnarray}
\label{E_absorbed}
  E&\approx& 2\pi \, \epsilon_r \, R_r \Delta R_r \Delta L_r \\
    &=& 7.53 \cdot 10^{-2} \, R_r \Delta R_r \Delta L_r \,
        \frac{J}{\mu \text{m}^3} \, . \nonumber
\end{eqnarray}
With the help of (\ref{E_absorbed}) the required radius $R_r$ is
approximately given by
\begin{eqnarray}
R_r&\approx&\frac{E}{2\pi \, \epsilon_r \, \Delta R_r \, \Delta L_r} \\
  &\approx& 13.2 \, \frac{E}{\Delta R_r \Delta L_r} \,
        \frac{\mu \text{m}^3}{\text{J}} \approx 2000
     \, \mu \text{m} \, , \nonumber
\end{eqnarray}
where it has been assumed that $\Delta R_r \approx 500 \, \mu
\text{m}$, $\Delta L_r \approx 30 \, \mu \text{m}$, and $E \approx 4
\cdot 10^6 \, \text{J}$ hold. The number of rods $N$ needed to absorb
the laser energy $E $ given $\epsilon_r$, $R_r$, $\Delta R_r$, and
$A_r$ is approximately
\begin{eqnarray}
N&\approx&\frac{2 \pi \, R_r \Delta R_r}{A_r} = \frac{E}{\epsilon_r
           \, \Delta L_r \, A_r} \approx 5 \cdot 10^7 \, ,
\end{eqnarray}
where $E \approx 4 \cdot 10^6 \, \text{J}$, $A_r \approx 0.25 \,
\mu \text{m}^2$, and $\Delta L _r= 30 \, \mu \text{m}$ have been
taken. Since the single effective rod can only withstand the laser
radiation for the time $\tau$ the required laser power $P_L$ is
approximately
\begin{eqnarray}
P_L&\approx& \frac{E}{\tau} \approx 8 \cdot 10^{19} \, \text{W} \, ,
\end{eqnarray}
where $E \approx 4 \cdot 10^6 \, \text{J}$ and $\tau \approx 50 \,
\text{fs}$ have been assumed. The required laser power $P_L$ can be
delivered by independent beamlines. If a single beamline is capable of
delivering $1 \, \text{kJ}$ and $2 \cdot 10^{16} \, \text{W}$ about
$4000$ independent beamlines are required.

The uniformity of laser energy deposition in the fuel is contingent
upon how the structured foams are distributed within the high-Z
hohlraum. It's important to emphasize that the option exists to
physically isolate the structured foams from the fuel, if necessary,
to mitigate the introduction of high-Z contaminants into the fuel,
while the structured foams can also be embedded into the fuel if
required.

It's worth highlighting that structured foams can be deliberately
designed to effectively convert the majority of incident laser energy
into electrons, ions, and radiation thereby generating significantly
elevated radiation energy densities. The subsequent reasoning serves
to illustrate this point.

In the absence of structured foams, achieving adequate radiation
homogeneity within a hohlraum necessitates multiple low intensity
reflections of laser radiation distributed over a larger
area. Consequently, the volume of such a hohlraum must be considerably
larger than the fuel pellet volume and in addition can only be
irradiated over an extended period of time. Hence, the energy density
in such a hohlraum is low. This scenario contrasts with the case of
structured foams filling a volume, wherein incident laser radiation
can be absorbed over the much smaller volume $R_r \Delta R_r \Delta
L_r$ in a fraction of the time leading to much higher energy densities
that are available almost instantaneously.

This distinctive characteristic translates to the fact that for
structured foams, at any given radiation temperature, the requisite
hohlraum volume is roughly equivalent to the fuel volume. This, in
turn, results in minimal wastage of laser energy for filling
expansive hohlraum volumes with surplus radiation energy. In addition,
the laser energy is deployed almost instantly.

\section{Conclusions}\label{sum}
In this paper, we introduce a novel direct drive fast heating
technology for reactive mixed $\ce{pBDT}$ fuels and simple low gain
fusion reactor designs based on them, which can reach $Q_T >1$ with
$\text{MJ}$ level heating energy. Simple low gain fusion reactors might
be building blocks of secondary applications. Specifically, they are
useful for the development of required laser and traget technologies
and are powerful neutron sources.

The heating technology involves the use of nano-structured
accelerators embedded within the fuel powered by ultra-short high
contrast laser pulses. The embedded nano-accelerators represent
embedded structured foams. These foams are capable of absorbing
$\text{MJ}$ level laser energy on sub-picosecond time scales, while
the energy deposition in the mixed fuel requires a couple of
picoseconds. The structured foams in combination with modern short
pulse lasers enable ultra-powerful ultra-fast direct drive fuel
heating avoiding parametric instabilities. The structured foams can
be used for constructing a new class of ultra-efficient utlra-high
energy density hohlraums.

By employing a simple numerical model, the impact of in-situ fusion
energy feedback and effective fuel confinement can be
analyzed. In-situ fusion energy feedback significantly reduces the
required scaled initial energy $E_i \, \rho_p/L$. According to
Fig. \ref{energies} the numerical model predicts that an initial
energy of approximately $3.4 \, \text{MJ}$ is sufficient to achieve a
fuel yield $Q_F >1$ for the simple reactor concept discussed here. It
is important to note that the numerical model can be adapted to fit
more sophisticated simulations with a community code with the help the
effective parameters $a, b, c$ for a given set of initial temperatures
$kT_e$ and $kT_i$, and initial density-range product $\rho_p R$, and
relative fractions of the fuel number densities involved.

It is important to note that energy production requires a target gain
$Q_T = \eta \, Q_F \approx 30 - 50$, as discussed in
\cite{perkins2021}, which is much larger than the one predicted for
the parameters investigated in the present paper.

\section{Acknowledgements}
The present work has been motivated and funded by Marvel Fusion
GmbH.

\bibliographystyle{elsarticle-num} 
\bibliography{literatur_eqn_motion}

\end{document}